# Adaptive and Context-Aware Volumetric Printing

Sammy Florczak[1,2], Gabriel Größbacher[1], Davide Ribezzi[1], Alessia Longoni[1], Marième Gueye[1], Estée Grandidier[1], Jos Malda[1,2], Riccardo Levato[1,2]*

[1] Department of Orthopaedics, University Medical Centre Utrecht, Utrecht University; Utrecht, 3584 CX, The Netherlands.

[2] Department of Clinical Sciences, Faculty of Veterinary Medicine, Utrecht University; Utrecht, 3584 CT, the Netherlands.

* Corresponding author. Email: r.levato@uu.nl

**Abstract:**

We introduce <u>G</u>ene<u>r</u>ative, <u>A</u>daptive, <u>C</u>ontext-Awar<u>e</u> 3D Printing (GRACE), a novel approach combining 3D imaging, computer vision, and parametric modelling to create tailored, context-aware geometries using volumetric additive manufacturing. GRACE rapidly and automatically generates complex structures capable of conforming directly around features ranging from cellular to macroscopic scales with minimal user intervention. We demonstrate its versatility in applications ranging from synthetic objects to biofabrication, including adaptive vascular-like geometries around cell-laden bioinks, resulting in improved functionality. GRACE also enables precise alignment of sequential prints, in addition to the detection and overprinting of opaque surfaces through shadow correction. Compatible with various printing modalities, GRACE transcends traditional additive manufacturing limitations, opening new avenues in tissue engineering and regenerative medicine.

## Introduction

The additive manufacturing workflow, originally conceived over forty years ago, has largely remained unchanged. While three-dimensional (3D) printing plays a crucial role across various applications in custom-made medical devices, microfluidics, bioengineered tissues, as well as in driving innovation in the automotive and aerospace sectors[1,2,3,4], the printing process always begins with users defining the desired part via computer-aided design (CAD) software, which is then translated to the printer, and finally fabricated either layer-by-layer, or using volumetric methods[5,6]. Further research on embedded sensors and feedback loops is mainly aiming to improve automation and is making significant strides towards performing in-line quality control of the printed objects[7,8,9,10]. Yet, 3D printers remain primarily tools that passively execute a command while being agnostic to the composition and nature of the environment in which the printing process is taking place.

Endowing printers with the capacity to capture and automatically respond to contextual cues would open new avenues in many high-end applications, including soft robotics, hierarchical composites, and the bioprinting of living cells and human tissues. In fact, the functionality of such systems is intimately linked to both their hierarchical architecture and relative positioning of their many individual components (i.e., particles, fibres, living cells)[11,12], whose precise patterning within printed objects cannot be fully controlled. Recent advances in computer vision and artificial intelligence are poised to revolutionize this approach. At the same time, the advent of volumetric printing (VP), including tomographic volumetric additive manufacturing (VAM), has enabled the extremely fast production of large parts, with virtually unconstrained design freedom. Visible light fields are used to polymerize photo-responsive resins without the need for layer-by-layer fabrication. Due to its contactless nature, volumetric printing excels at overprinting - that being, non-invasively printing onto or across existing objects - even those produced with other techniques[13]. This includes producing multi-material structures[14], and safely sculpting resins containing fragile living cells and organoids[5]. Such attributes position VP as an ideal demonstration platform for new fabrication pipelines that would be challenging to implement using traditional manufacturing techniques.

In this study, we report a new technique to endow 3D printers with the ability to "see" and perceive the composition (chemical and architectural) of the printable material, and to take autonomous, informed decisions on which geometries to print. This novel workflow allows the printer to automatically produce, within seconds, volumetric, generative designs that adapt and conform to the contents of the resins, of which the printer is aware. This approach is compatible with a wide range of fabrication technologies and materials, including complex composites, and living cells with bio-friendly hydrogels, unlocking a new generation of applications in tailor-made tissue-mimetic materials and data-driven rapid prototyping.

## 3D Printing Guided by Volumetric Imaging

We termed this workflow: <u>G</u>ene<u>r</u>ative, <u>A</u>daptive, <u>C</u>ontext-Awar<u>e</u> 3D Printing (GRACE). We first demonstrated GRACE by empowering volumetric printing with light sheet imaging - this being the "eyes" of the printer. The light sheet rapidly maps the printing volume in 3D to extract positional, morphometric, and spectral (i.e., fluorescence) information from the contents of the vial. This serves as input for multi-parametric modelling algorithms – acting as the "brain" – to generate precisely targeted geometry for printing. While GRACE printing is relevant for many high-end fabrication processes, we first showcased its key capabilities in the context of bioprinting, using cell-laden hydrogels as resin platforms. In living organs, key tissue components develop to adapt to the needs of the resident cells. For instance, blood vessels grow into intricate networks to reach each individual cell ensuring access to nutrients. Currently, printing technologies cannot fully recapitulate this process, as cells are randomly or homogeneously distributed within a printed hydrogel. With GRACE, we demonstrated on-the-fly generation of 3D models to create positive and negative features, including targeted capillary-like vessels that can precisely reach cells, cell clusters and organoids of interest, resulting in improved viability and functionality of the bioprinted cells. We further demonstrated the production of interconnected geometries, and the precise encapsulation of various particulate elements embedded within the resin. Moreover, we enabled the automated alignment of new prints onto pre-existing ones, permitting the generation of multi-component tissues with native-like cell density. Light sheet imaging also enabled the mapping of opaque features, allowing for mitigation of shadowing artifacts to improve (over)printing quality. Finally, we also demonstrated the compatibility of GRACE with multiple 3D printing modalities, including embedded extrusion printing in suspension baths [15], and filamented light projection fabrication (FLight)[16].

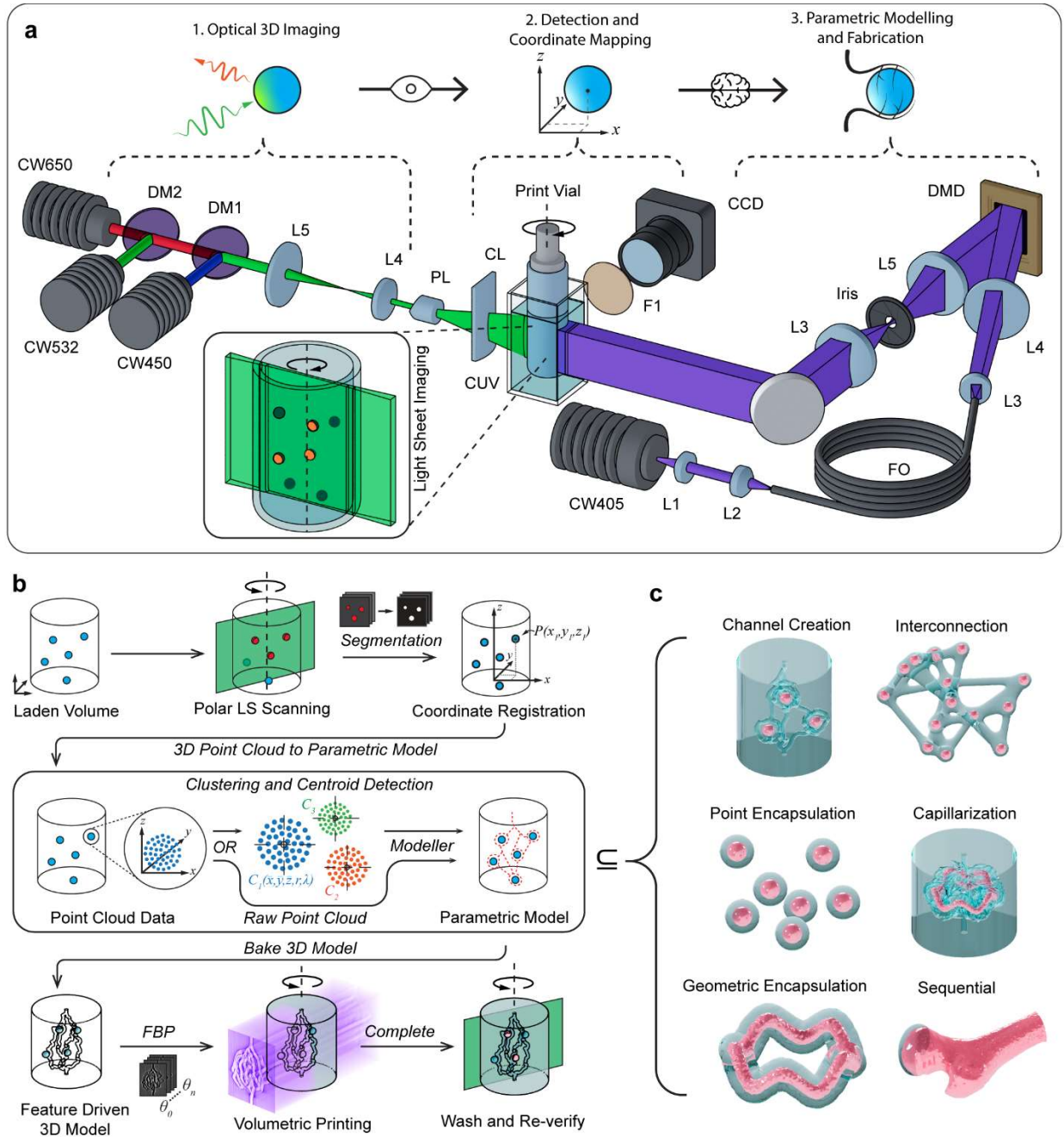

*Fig. 1 | Experimental setup and workflow for GRACE printing. a)* Optical schematic of experimental device, with the printing path in violet and lightsheet imaging path in green. The printing path comprises collimation and coupling optics (L1, L2), square core multimode-fibre (FO), DMD, and 4f-relay (L3, Iris, L5). The lightsheet path comprises dichroic beam combining optics (DM1,2), beam reduction optics (L5, L4), 30° Powell lens (PL), and cylindrical lens (CL) for focusing the light sheet. *b)* Flowchart of the GRACE methodology, demonstrating the process of lightsheet scanning feature-laden volume, feature

*segmentation and coordinate registration, using either the raw data or clustering algorithms to then generate a parametric model, and print. **c)** A subset of various parametric geometries (both positive and negative) generated using various point-like and bulk scanned features (displayed in red).*

To perform GRACE, we have designed an experimental device comprising two main optical paths: a custom made tomographic volumetric printer; and a light sheet microscopy path (Fig. 1a). We chose polar light sheet microscopy as the primary scanning modality due to its capacity for rapid, large-scale imaging, and its ease of incorporation within the setup, requiring no modification to the printing path. Light sheet generation in our implementation could be achieved through two complementary methods: i) by digitally encoding a light sheet pattern (i.e. a single column of activated pixels) onto the digital micromirror device (DMD), or ii) utilizing an external laser source with dedicated optics. Although the former required no additional optics or excitation sources, for most of our experiments, we opted for an external light sheet configuration to maximize power and signal-to-noise ratio during scanning, especially as fluorescence was utilized as the primary contrast agent. A refractive index matching water-filled bath was also present for minimizing refractive errors while printing and imaging.

To enable the 3D mapping and registration of features embedded within the printing volume we developed a protocol which synergized the motion hardware with the light sheet imaging and printing systems (see Fig. 1b). Upon initial homing of the vial, the light sheet illuminates the sample, acquiring images that will serve as the input of the computer vision routines. The resultant fluorescent emission of the optical section is then captured by the camera at many angular increments to obtain a dense set of polar optical sections through the axis of the vial. This can then be repeated multiple times for different fluorescence channels. To extract useful information from the polar stack, thresholding and segmentation is performed to isolate features-of-interest from the background. The extracted data is then assigned corresponding per-pixel Cartesian coordinates, producing a volumetric 3D point cloud in print-volume coordinate space. Depending on the scanned features and intended model, the raw data can then either be passed directly to parametric modelling software or undergo additional processing in the form of cluster detection. For this, we used Density-Based Spatial Clustering of Applications with Noise (DBSCAN), chosen for its efficiency with large datasets, ability to identify arbitrarily shaped clusters, automatic outlier exclusion, and lack of requirement for pre-defining the cluster count [17]. Clustering detection was applied where numerical indexing of individual features was required or where centroid coordinates were preferred, as is the case with symmetrical features such as particles, organoids, or microspheres. This workflow provided robust feature detection in complex, heterogeneous samples, with the resultant coordinate data passed to the parametric modelling software to automatically generate bespoke geometries targeted around those scanned features. A subset of these possible context-driven parametric geometries is illustrated in Figure 1C.

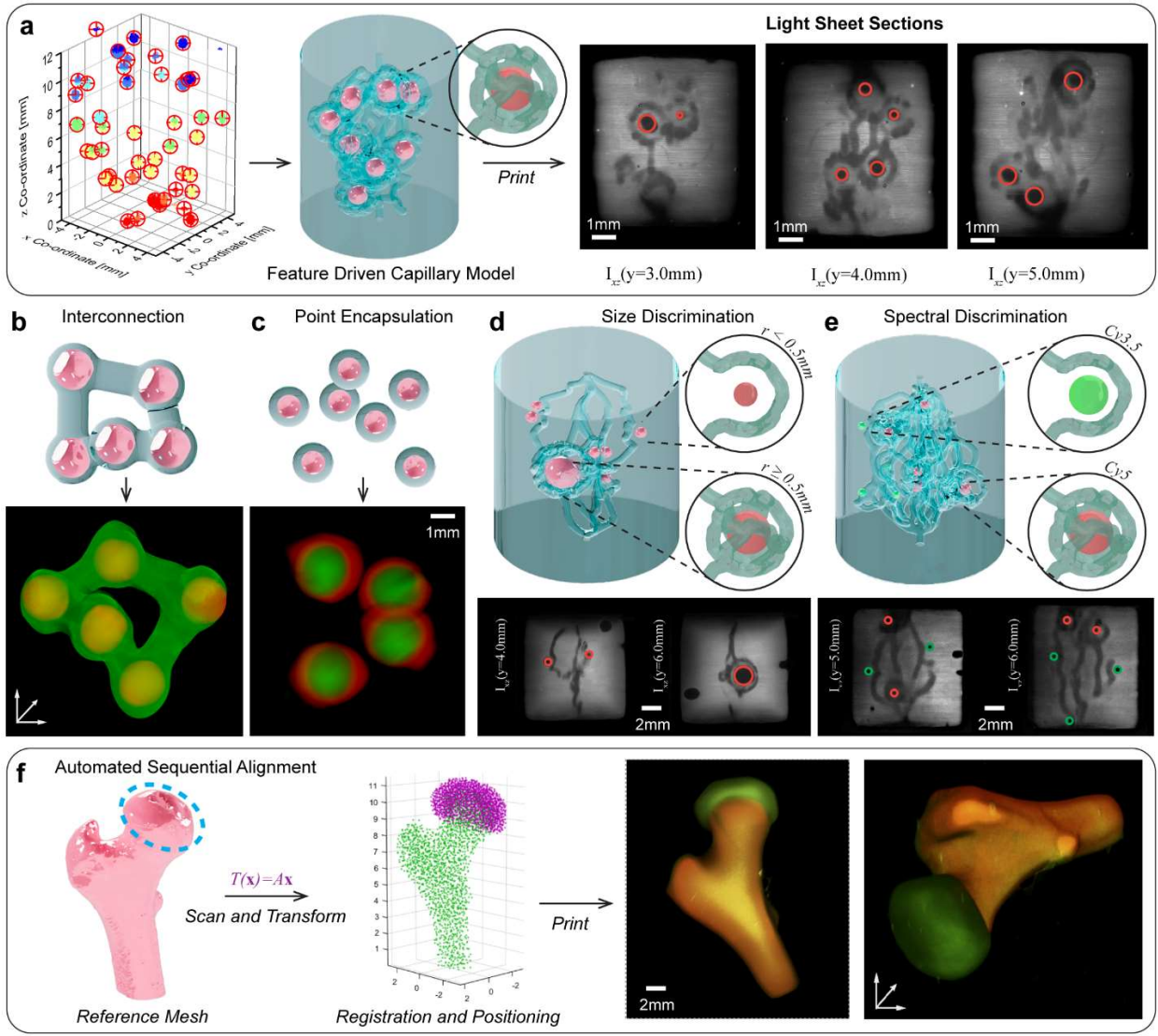

***Fig. 2 | GRACE allows printing adaptive and feature-driven prints with complex geometries.*** *A showcase of GRACE printing by generating targeted features around randomly distributed fluorescently stained alginate spheres detected within the resin.* ***a)*** *A capillary-ball channel architecture is generated around the alginate spheres, showing the process from acquiring the raw data, to generating the model, and finally printing and imaging the resultant construct. Lightsheet micrographs show sections at various positions, with the alginate spheres indicated by red circles.* ***b)*** *Interconnection of randomly distributed alginate spheres with printed struts, showing the target geometry and resultant reconstruction in 3D after printing.* ***c)*** *Encapsulation; with corresponding lightsheet 3D reconstruction post-printing indicating presence of spheres. Parametric discrimination of generated features based on* ***d)*** *Size or* ***e)*** *Spectral emission. Here, different populations receive either a single grazing channel or a more complex capillary-ball.* ***f)*** *Automated alignment of cartilage model to femoral head in a two-part sequential print.*

## Context-Driven Parametric Structures Using GRACE

To demonstrate the capabilities of the GRACE workflow, we successfully printed a series of complex, context-driven parametric architectures. As a testing platform, we produced fluorescently stained spherical alginate particles of various sizes between 0.3mm to 1.8mm. These were suspended and mixed into a gelatin methacryloyl resin (GelMA, 10% w/v), with lithium phenyl (2,4,6-trimethylbenzoyl) phosphinate (LAP, at 0.1% w/v) serving as a photoinitiator. Using only spatial information with clustering detection to determine the centroids of the alginate particles, we first successfully demonstrated context-aware prints with vascular-like channels (450±20μm) targeted around individual spheres with a pre-defined 300μm offset from the surface (Fig. 2a). We also fabricated strut-like interconnections between multiple particles (Fig. 2b), and performed single-layer encapsulation of individual features (Fig. 2c). Building upon these results, we further leveraged morphometric and spectral data to discriminate between different sphere populations, allowing for more sophisticated architectures to be generated. To demonstrate size-based discrimination, we combined alginate particles of different radii (ranging between 0.2 mm and 1.8mm) within the same volume. Cluster detection was then used to index each point cloud corresponding to each particle, allowing individual radii to be measured. The resulting index and radii table for the scanned features was then utilized by our parametric model to generate conditional geometry defined by a size threshold, such that spheres with $r < 0.5$ mm received a single grazing channel, while those with $r \geq 0.5$ mm were enclosed by a complex capillary ball (Fig. 2d). Extending this concept further, we applied a similar conditional approach to discriminate between differently stained spheres - successfully leveraging spectral information to guide the parametric model in generating distinct architectures based on the fluorescence signatures of Cy3.5 and Cy5. In this case, the model generated solitary grazing channels around alginate particles stained with Cy3.5, whilst convoluted capillary balls were generated around Cy5 stained particles (Fig. 2e). Excluding the time required for vial loading, these examples were completed in a streamlined manner; with scanning, feature isolation, model generation, and printing being completed within approximately 4 minutes (per spectral channel). This underscores the efficiency and compatibility with the rapid, high-throughput capabilities characteristic of VBP. For a detailed breakdown of processing times for this, and subsequent modalities of GRACE, refer to Supplementary Table 1.

Finally, we demonstrated the capability of the GRACE workflow for sequential, multi-step fabrication through the implementation of an automatic alignment feature. This eliminated the need for manual repositioning of sequential prints within the resin - a process[6,18] that is tedious, error-prone, and prohibitively slow. Instead, our system automatically detects and aligns with previously printed structures, enabling the precise positioning of subsequent prints relative to existing ones. This automated approach streamlines the fabrication workflow to allow for creation of complex, multi-layered or hierarchical structures with high repeatability. We successfully demonstrated this by first printing a model of a human femur, washing it, and resuspending it in GelMA at a random orientation. We then employed an iterative closest point (ICP) algorithm to determine the optimal rigid transformation that aligns a pre-existing reference model (featuring correctly positioned cartilage) to the scanned point cloud of the femur. With the alignment computation (post-scanning) taking <5 seconds to perform on our system (Extended Data Table 1), the transformed cartilage model was printed directly onto the femoral head, resulting in a multi-component sequentially printed construct (Fig. 2f) with correct relative positioning of its components.

Our experiments demonstrated how GRACE can adapt to various embedded objects and generate precise, functional geometries with minimal user input after the initial parametric model definition. This level of automation and adaptability would be impractical, if not impossible, to achieve through manual positioning and 3D modelling. While these examples demonstrate the versatility of our system, they represent only a fraction of its capabilities, with the workflow able to accommodate virtually any array of parametric models and data streams for generating context-driven architectures.

## Image-guided Printing across Opaque Features

Leveraging the adaptive capabilities of GRACE, we next addressed the presence of shadowing artifacts introduced by light-absorbing features within the print volume - a common challenge in light-based printing. It is well established that occlusions present within the projection path result in poor reconstruction quality, compromising the dimensional accuracy and homogeneity of the printed structures when performing overprinting, especially for VAM[6,19,20]. We successfully utilized our light sheet imaging modality as a profilometer, using the reflected signal to map the surfaces of occluding structures within the print volume and thus enabling us to mitigate these artifacts (Fig. 3a). We utilized Object-space optimization of tomographic reconstructions (OSMO) to iteratively optimize our tomograms for the presence of these occlusions[21,22]. This was demonstrated in two ways. First, we used an opaque, polymeric occlusion produced via stereolithography, consisting of ten 1mm diameter vertical pillars to provide a reproducible occlusive feature within our build volume. The quantity and spatial distribution of pillars was based on a Monte Carlo optimization by minimizing the Bhattacharyya coefficient (for contrast) and maximizing the Jaccard similarity index - both derived from the OSMO corrected reconstructions (Fig. 3b). The resulting occlusion phantom (Fig. 3c) was embedded into our resin and scanned, allowing us to parametrically generate the representative occlusive geometry based on the pillar centroids and angles. Upon optimizing and printing the target geometry (8-toothed cog model, Fig. 3b), the shadow-corrected structure qualitatively demonstrated a visible improvement in print quality compared to uncorrected output, with the latter showing regions of both under- and over-crosslinking within the same print (Fig. 3d). This indicated that there would exist no possible dosage condition for which an accurate output could be attained without shadow correction. Meanwhile, the corrected output displayed a more homogenous crosslinking behaviour, with finer details retained across the entire construct.

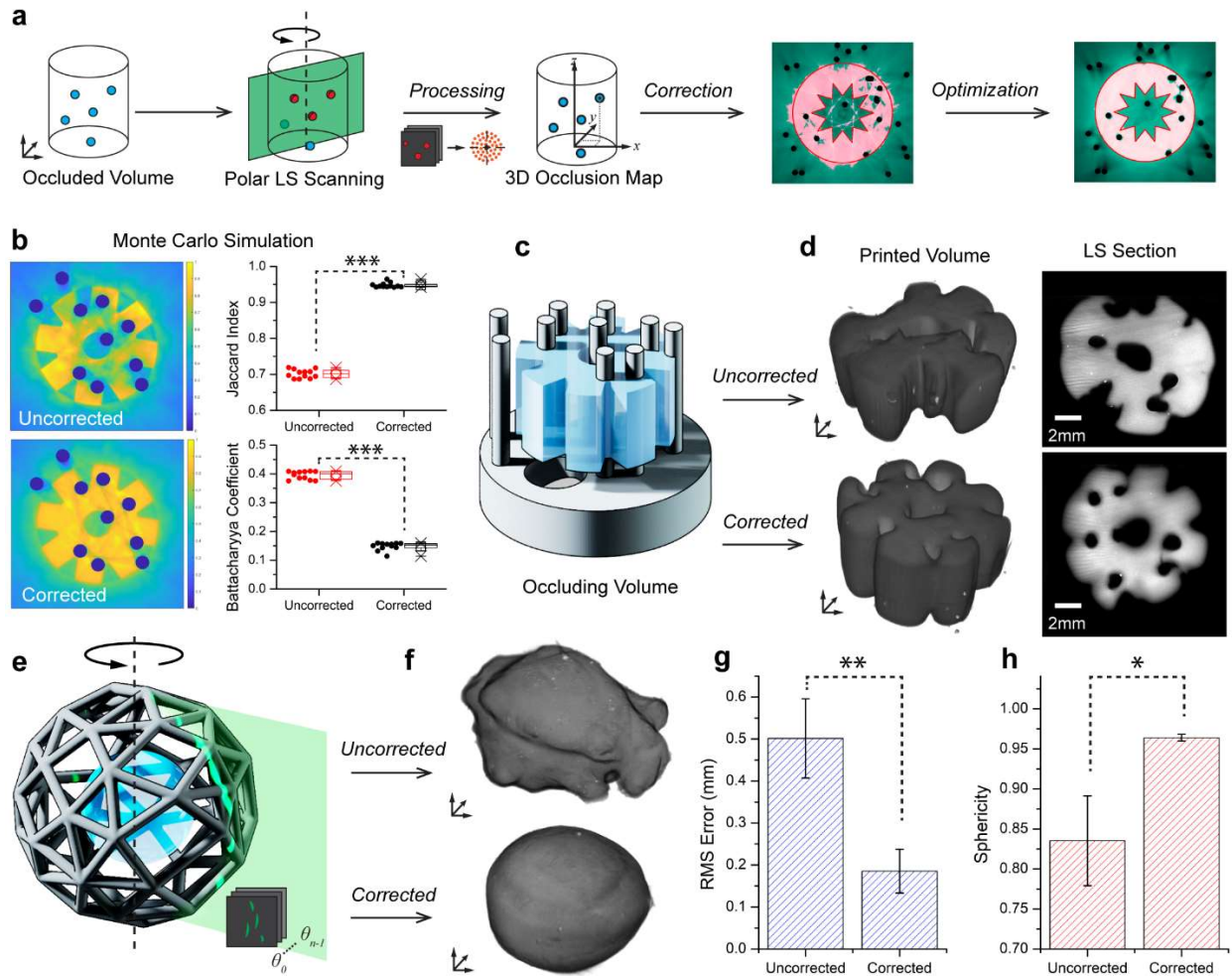

*Fig. 3 | Light sheet mapping of occluding structures and shadow correction. a)* A flowchart demonstrating the process of scanning, mapping, and correcting for the presence of occlusions. *b)* OSMO-based reconstruction of a cog-like target model without and with shadow correction when influenced by 10 randomly distributed pillar-like occlusions. Jaccard and Bhattacharya Coefficients demonstrate the relative improvements attained by the correction (n=12, p<0.001). *c)* Rendering showing the SLA printed pillar occlusions, and cog-like target geometry to be volumetrically printed around them. *d)* 3D lightsheet reconstruction of corrected and uncorrected prints, in addition to a single optical section of the cross-section of each. *e)* Render showing the ball-in-cage model, with spherical target geometry inside. *f)* Lightsheet reconstruction of the resultant prints following the destructive removal of the occluding cage following printing. *g)* Root-mean-square error of the printed part. A lower value indicates less deviation from the target geometry (n=3, p < 0.01). *h)* Sphericity of each printed sample. A higher value indicates the sample is more spherical (n=3, p < 0.05).

This is further corroborated by the Monte Carlo simulation, which indicated improvements would be achieved in both contrast - evidenced by a reduction in the Bhattacharyya coefficient from 0.39±0.01 to 0.15±0.01 - and in the similarity index, which increased from 0.70±0.01 to 0.945±0.007 after correction.

To demonstrate applicability to more complex continuous occlusions, we further extended this approach, aiming to print a trapped ball-in-cage model (Fig. 3e). This presented a more challenging scenario, with the cage structure creating multidirectional occlusions around the central sphere. Unlike our previous approach, where we parametrically generated the representative occluding surfaces using the scanned point cloud data, here we instead combined our auto-alignment workflow to precisely match the *a priori* reference model to the scanned point cloud, thus aligning, rather than generating the occluding geometry. As before, the aligned mesh was used as the occlusion input for the OSMO-based optimization to generate a corrected set of tomograms. The output (Fig. 3f) demonstrated a statistically significant improvement in the shadow-corrected model compared to the uncorrected print, as evidenced by both a reduction in surface root-mean-square error (RMSE) from $0.50\pm0.09\mu m$ to $0.18\pm0.05\mu m$ and an increase in sphericity from $0.83\pm0.06\mu m$ to $0.965\pm0.006\mu m$, for the uncorrected and corrected prints respectively (Fig. 3g, h).

These experiments highlight the suitability of GRACE to be used not only for creating context-aware geometries, but also for mapping occlusive features within the printing volume and mitigating their influence.

## Bioprinting with GRACE

Having demonstrated GRACE's capabilities with synthetic embedded objects, we next explored its potential for biofabrication applications. This rapidly growing field is aimed towards producing highly structured engineered tissues for regenerative medicine, and for developing *in vitro* models for personalized medicine. Specifically, we aimed to fabricate adaptive geometries around living cells and organoids, showcasing *i)* the production and functionality of customized vessels, *ii)* the generation of automatically aligned multi-tissue prints, and *iii)* the compatibility of GRACE with printing techniques other than volumetric printing. As a first demonstrator - inspired by how *in vivo* blood vessels grow to reach cells and provide them with necessary nutrients - we assessed our workflow to automatically generate adaptive vessel-like networks optimized around dense cellular structures. Notable efforts in the field of bioprinting have focused on producing convoluted channel networks to nurture engineered tissues[23,24,25]. However, none of these approaches can print pre-defined channels capable of reaching every cell or organoid within the construct, as current printers lack the ability to adapt their designs to the specific needs of each cellular structure within the printing volume.

Herein, we successfully combined GRACE with Embedded Extrusion-Volumetric Printing (EmVP)[26] to generate adaptive vascular-like architectures around toruses densely laden with insulin-secreting, engineered pancreatic cells (iβ-cells, $5.0\times10^7$ $mL^{-1}$) at tissue-relevant cell densities[27]. These toruses were extruded into a GelMA-based bioresin (Fig. 4a).

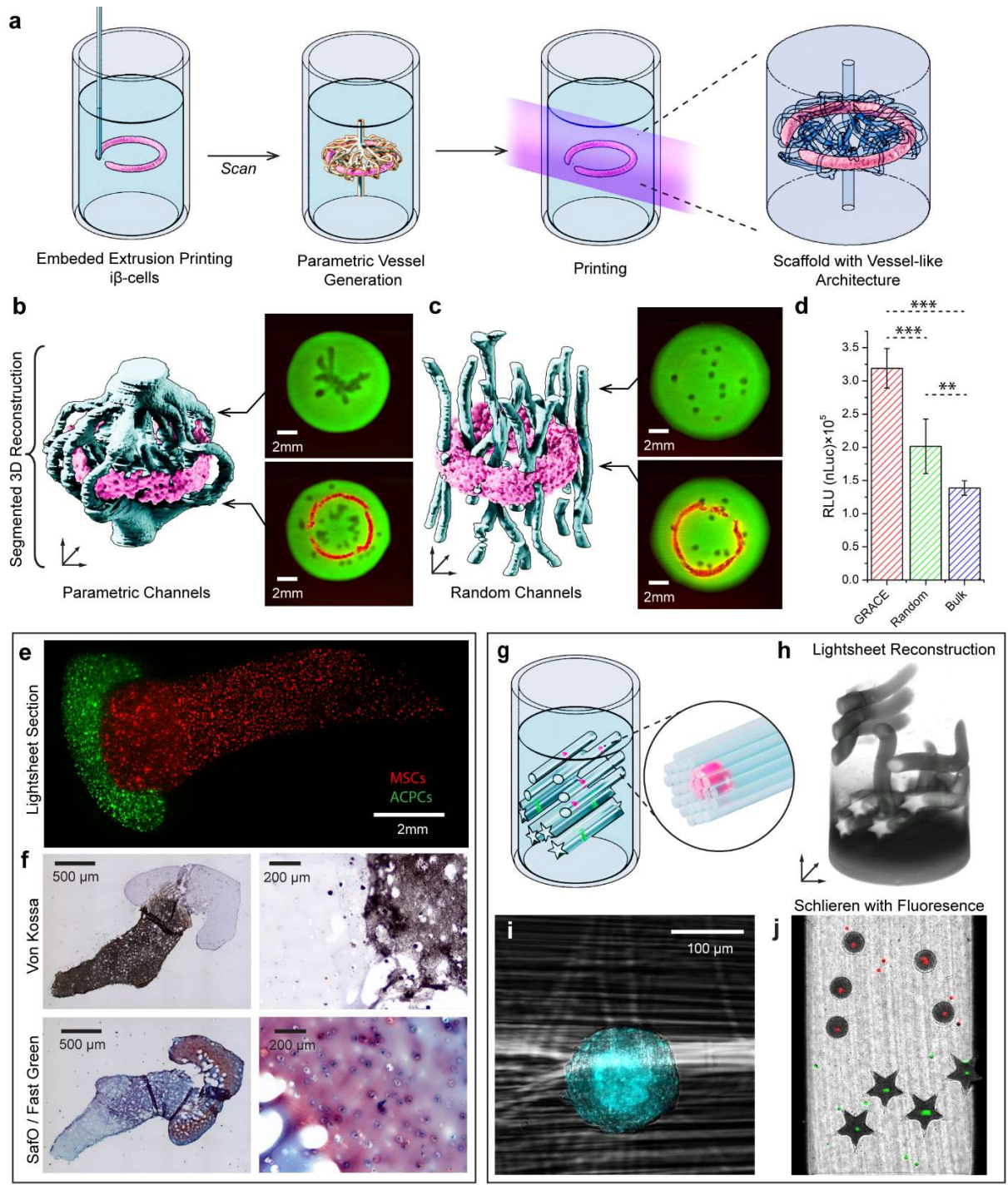

*Fig. 4 | **Bioprinting of functional living tissues with cell-location driven features enabled by GRACE. a)** Diagram of experiment process depicting the scanning of the volume, feature-driven model generation around the torus, and printing of the scaffold. **b,c)** 3D segmentations of the resultant structures printed around the extruded toruses, both targeted, and random, also showing light sheet micrographs of the scaffold cross-sections at two planes. **d)** Graph showing the amount of insulin stored and released by iβ-cells via the bioluminescent reporter nanoLuc (n≥3, \*\*p<0.01, \*\*\*p<0.001). **e)** A light sheet section of the construct immediately following printing with the automatically aligned femur-cartilage model showing distinct osteal and chondral regions. **f)** The histological sections following 4 weeks of maturation highlight the presence of a mineralized compartment (Von Kossa positive, brown/black staining) and of*

*glycosaminoglycans rich cartilaginous tissue (SafO, red staining).* ***g)*** *3D model of FLight construct generated with GRACE after scanning.* ***h)*** *Resultant light sheet 3D reconstruction after printing and washing.* ***i)*** *Filamentous construct is visible surrounding the spheroid.* ***j)*** *Discrimination of spheroids is evident in the Schlieren image (captured immediately after printing) with fluorescence overlay.*

Following scanning, we generated and printed the scaffolds around the extruded bioink toruses, with channel diameters of 1mm at the inlets tapering to 0.4μm at the torus surface, maintaining a fixed surface area of ~180±10mm² with a 300μm offset from the torus (Fig. 4b). To evaluate the efficacy of this approach, we compared these adaptive structures to two controls: randomly generated channels of the same surface area (Fig. 4c), and a bulk structure without channels. Following 24h culture of the sample in dynamic culture conditions, we observed a significant increase in proinsulin secreted in the supernatant of the adaptive structures compared to both the random non-targeted and bulk structures (Fig. 4d). This result suggests superior mass transport within the adaptive geometries when compared to randomly distributing channels throughout the construct, highlighting the advantage of our context-aware approach.

Next, we leveraged our previously described automated alignment workflow to sequentially fabricate a two-component cell-laden bone and cartilage model. Here we prepared two resins containing articular cartilage-derived progenitor cells (ACPCs) and bone-marrow derived mesenchymal stem cells (MSCs) encapsulated in GelMA at cell concentrations of $1.0 \times 10^7$ mL$^{-1}$ and $5.0 \times 10^6$ mL$^{-1}$ respectively. Three femur models were first printed within the MSC-laden GelMA, washed, then placed within the printing vat filled with the ACPC-laden bioresin. The MSC component was scanned to determine the orientation and position of the femur, and to enable the cartilage to be automatically positioned onto the model. The parametric model allowed for on-the-fly adjustment to the cartilage thickness. The cartilage phase was then printed to form a layer around the femoral head, resulting in a multicellular construct that mimics the native osteochondral tissue positioning (Fig. 4e). Cells remained viable and persisted in their intended compartment (chondral or osteal) over 4 weeks of culture. During this period, ACPCs and MSCs synthesized cartilage and mineralized bone matrix components, respectively. Histological analysis confirmed the correct formation of a multi-component tissue structure (Fig. 4f).

Finally, we demonstrated the versatility of our technique by using an alternative printing modality: filamented light (FLight) biofabrication, a vat polymerization technique that generates patterned structures composed of multi-centimetre long, aligned microfilaments[16]. Here, two differently stained populations of spheroids composed of aggregated MSCs were prepared (DiD or DiO) and suspended within a 10% GelMA+LAP resin. The coordinates of each population were determined, and a parametric model was written to assign each group a unique shape (either stars or circles for DiD or DiO stained spheroids, respectively, Fig. 4g). Printing was then performed around a subset of each spheroid population (to avoid overcrowding). This resulted in the formation of elongated filamentous constructs spanning the width of the print volume, with the spheroids precisely positioned through the central axis of each star or circle FLight structure (Fig. 4h,i,j). This demonstration showcases the ability of GRACE to adapt to different printing techniques while maintaining its core capability of generating context-aware, population-specific geometries.

## Discussion and Outlook

This study introduced GRACE, an innovative workflow that leverages the unique attributes of volumetric printing to enable the fabrication of context-aware geometries. By integrating light sheet microscopy, computer vision algorithms, and parametric modelling, GRACE allows us to detect and respond to features across multiple scales, from individual cells to macroscopic structures. Our results illustrate the ability of the system to rapidly generate and fabricate complex geometries that dynamically adapt to arbitrarily distributed features within the print volume. This level of automated, context-driven fabrication would be prohibitively time-consuming and impractical to achieve through manual design processes. GRACE operates with minimal user intervention, requiring only experiment-specific adjustment of the parametric models, thus significantly streamlining the fabrication workflow while simultaneously expanding the complexity and functionality of achievable structures. The workflow's versatility extends beyond its current implementation, as the concept can be directly integrated into other volumetric or layer-wise printing approaches. For instance, it could be seamlessly incorporated into xolography, which inherently utilizes light sheet optics[28]; multiphoton printing, acoustic-based volumetric printing[29,30], or extrusion printing in suspension baths[31]. Furthermore, the adaptability of GRACE allows for exploration of alternative imaging modalities, such as optical tomography (including schlieren, diffraction, coherence, or diffuse), or holographic approaches[32,33,34], for non-fluorescent imaging.

The GRACE workflow, as introduced in this work, also has direct implications for biofabrication, enabling new possibilities in several key areas. These include the creation of biomimetic scaffolds that can adapt to the spatial distribution of cells or organoids, and the fabrication of tissue constructs with highly controlled microenvironments, which are known as potent regulators of cell function, differentiation, and tissue maturation. GRACE also opens new avenues as a tool for adaptively modifying the properties of an object at any timepoint after printing. Such modifications could include spatial-selective grafting of chemical compounds and proteins[35,36], establishment of stiffness gradients[37] or modulation of viscoelasticity[38]. Altogether, these advancements underscore the versatility of GRACE and potential in both general additive manufacturing and bioprinting contexts, representing a paradigm shift in how printing is performed.

## Methods

### Tomographic volumetric printer

The custom-built volumetric printer employed in this study (Fig. 1a) utilised a 405nm laser source shaped into a flat-top intensity profile via coupling of the beam into a square core fibre (WF 70x70; CeramOptec, Germany). This shaped beam was collimated and used to illuminate a DMD (Hispeed v-7000; ViALUX, Germany), for which the resultant projection was Fourier filtered and imaged with 1:1 magnification using a *4f* relay onto the centre of the printing volume in a telecentric projection regime. During the printing process, vials were immersed within a refractive index compensating bath comprising a square water-filled quartz cuvette (OP36; eCuvettes, China). This was required to minimize refractive errors arising from the curvature of the vial surface, both for printing and feature scanning. To accommodate different print requirements, we used borosilicate glass vials (Readily3D, Switzerland) with internal diameters from 8.8 mm to 15.3 mm, selected according to the specific model parameters and target print volumes.

**Light sheet imaging module**

The light sheet imaging path (Fig. 1a) employed three 40-50mW diode lasers operating at 450nm (RLDD450-40-5; Roithner, Austria), 532nm (Roithner RLDD532-50-4) and 650nm (Roithner RLDH650M-40-5) to cover a broad range of commonly used fluorophores. The beams were combined with dichroic mirrors, shaped into a flat-top fan profile with a 30° Powell lens (43-473; Edmund Optics, USA), then focused to the vial centre using an f=50mm cylindrical lens. Two scanning regimes were employed: a half-sweep mode covering $\Theta_{total}= \pi$ (500 polar sections) and a $\Theta_{total}=2\pi$ sweep mode (1000 polar sections). The latter was preferentially used for samples exhibiting occlusive or highly scattering samples, as this provided bi-directional illumination of the features. Image acquisition for feature scanning/registration was performed using a monochromatic camera (Alvium 1800 U-240m; Allied Vision, USA) in conjunction with an f=50mm C-Mount lens (MVL50M23; Thorlabs, USA). Exposure and gain parameters were manually set prior to each scanning session. A rotary filter mount containing bandpass filters for several commonly used fluorophores (GFP, Cy3.5, Cy5) was integrated post-objective to facilitate spectral selection of the emission signal while rejecting the backscattered laser line.

**Image processing and feature registration for GRACE**

A MATLAB script was prepared to perform several functionalities key to the GRACE workflow. This process (shown in Fig. 1b) was responsible for the following: i) the initialisation and synchronisation of hardware (imaging, light sheet, and printing) and software parameters; ii) the acquisition and processing of polar light sheet image stacks; iii) the isolation and registration of features-of-interest from the background; iv) the conversion of image data to useable 3D coordinate data; v) the processing of this data (e.g. using clustering detection) to extract construct-specific information necessary for generating the desired construct, and; vi) outputting the said data for use with the parametric modelling software.

**Data-driven parametric models**

We employed off-the-shelf software Rhino3D (Robert McNeel & Associates, USA) in conjunction with its integrated Grasshopper visual programming environment. Grasshopper (GH) definitions were developed to create parametric models for each geometry type. These definitions performed three key tasks: i) importing and synchronizing with data files containing coordinates, radii, and other relevant data as described above; ii) utilizing this data to generate the desired parametric geometry; and iii) baking and exporting the final model as an STL file suitable for 3D printing. Our work focused on three broad types of bio-inspired geometries: perfusable vessel-like channels surrounding scanned features with an inlet and outlet, positive interconnected geometries, and targeted single-layer encapsulation. In several of these cases, while not strictly necessary, we also leveraged two freely available add-ons for GH. These were the DENDRO plugin[39], which facilitated convenient surface generation around point structures; and the ShortestWalk plugin[40], which makes use of the A* algorithm[41] to calculate the shortest walk in a network of paths. We note that similar functionalities can be provided using other plugins (either inbuilt or 3rd party), or by writing a custom script within GH.

**Synthesis of GelMA**

All chemicals were obtained from MilliporeSigma, and used without further purification or modification, unless stated otherwise. Gelatin Methacryloyl (GelMA) was synthesized as previously reported[42]. Briefly, 0.6 g of methacrylic anhydride were added per gram of gelatin (type A, from porcine skin, 10 w/v% in phosphate buffered saline, PBS), and left to react for 1 hour at 50 ˚C under constant stirring, to obtain a degree of methacryloyl substitution of 80%, as assessed via 1H-proton nuclear magnetic resonance (NMR; 400 MHz, Agilent 400 MR-NMR, Agilent technologies, USA). The resulting solution was dialyzed (MW cut-off = 12 kDa) against deionized water to remove the unreacted methacrylic anhydride. The purified macromer solution was sterile filtered (0.22 μm), freeze-dried and stored at -20 ˚C until used.

### Resin preparation and printing

Gelatin methacryloyl (GelMA)-based bioresinsupted with 0.1% w/v lithium phenyl-2,4,6-trimethylbenzoylphosphinate (LAP, Tokyo Chemical Industry) photoinitiator was used for experiments in this work. Prior to printing, resins were thermally gelled by immersing in ice-water for 10 minutes. Schlieren imaging was leveraged to observe the crosslinking process during printing, for which the process was manually ceased upon observing a sufficient accumulation of dosage, as evidenced by a rapid change in the refractive index of the resin. After printing, samples were washed with warm (37°) PBS to remove un-crosslinked material.

### Linear light sheet scanning of samples post printing

For all experiments in this paper, printed samples were imaged using a custom built, linearly swept light sheet fluorescence microscope. Samples were placed in a 22mm × 22mm square cuvette (OP36; eCuvettes, China) and immersed in PBS during imaging. Sections were captured at 35μm steps over the span of the entire sample.

### Preparation of alginate microparticles

Alginate microparticles were produced by dissolving sodium alginate (alginic acid sodium salt, MilliporeSigma) in deionized water at 2% w/v concentration. A crosslinking solution was prepared by dissolving $CaCl_2$ (Sigma-Aldrich) in DI water, reaching a final concentration of 100 mM. An in lab-made platform was purposely built for microparticles production, consisting of a coaxial needle system (RegenHu, Switzerland), where the inner needle was used for airflow and the outer for alginate flow. Two Inkredible bioprinters (Cellink, BICO Company, Sweden) were used to regulate the pressure for both the airflow and alginate flow rate. A collection bath, filled with 100 mM calcium chloride solution, was utilized for the collection and crosslinking of microparticles. The collection bath was placed below the coaxial needle and on top of a magnetic stirrer, which was used to homogeneously stir the solution in the collection bath. Once the setup was built, alginate hydrogel solutions with different fluorescent dyes (Cy-5 and Cy-3.5) were prepared and loaded into 3 mL cartridges, which were finally connected to the coaxial needle. The air pressure and flowrate were regulated to produce different sizes of microparticles. After generating the microparticles, they were strained and transferred into a 50 mL Falcon tube and washed using deionized water. A fluorescent microscope (Leica Microsystems SP8, Germany) was used for imaging and characterization of microparticle size.

### Using GRACE to detect and print around alginate particles

For all work involving alginate microparticles, resins were prepared with 10% GelMA w/v + 0.1% LAP w/v. The stained alginate particles were laden into the volume, and gently suspended by agitating the volume while being cooled in an ice bath until the resin thermally gelled. These were then scanned over $\Theta_{total} = \pi$, and cluster detection was used to determine the centroid coordinates of each alginate sphere (as described above in the pipeline for feature coordinate registration in GRACE). For prints involving two populations of stained alginate particles, this process was performed twice to obtain a separate coordinate dataset for each group. The corresponding adaptive models were then generated and printed according to the GRACE workflow.

**Auto-alignment of sequential prints**

Auto-alignment of sequential prints was accomplished using a combination of GH and MATLAB. In our work, this was demonstrated with a femur and cartilage model, though the same workflow can be performed to automatically align any two or more geometries sequentially. The process began by importing a reference femur model into GH, where a section of geometry from the femoral head was extracted and assigned an adjustable extruded thickness, allowing for customizable cartilage thickness. The relative positioning between the two models defined the target alignment between the sequential geometries *a priori*. To accomplish this alignment practically, we sought to determine the rigid transformation required to align the reference femur (and thus the relatively placed cartilage model) to the location of the arbitrarily located scanned femur within the vial. First, a stained femur (Cy5) was printed using 10% GelMA w/v + 0.1% LAP w/v, washed, and resuspended into new GelMA resin (stained with Cy3.5) at a random orientation. This print was then scanned, and the resulting volumetric point cloud data $P_{scan} = \{q_1, q_2 \ldots, q_m\}$ was exported. A dense array of random points $P_{Ref} = \{p_1, p_2 \ldots, p_n\}$, was then generated within the volume of the reference femur model then also exported to MATLAB. The reference femur point cloud ($P_{ref}$) was automatically aligned to the scanned femur ($P_{scan}$) using the iterative closest point (ICP) algorithm (MATLAB's 'pcregistericp' function). This generated a rigid transformation matrix $T$, such that when also applied to the reference cartilage geometry $G_{Ref}$, resulted in a rotation and translation of the cartilage component to its correctly aligned position over the scanned femoral head within the vial, such that $G'_{Aligned} = TG_{Ref}$. The transformation data was synchronized to the GH definition, thus generating a correctly placed cartilage geometry that was subsequently exported and printed. By using any reference models and assigning correct relative positions within GH, this process can be leveraged for any arbitrary alignment task, providing a versatile approach for complex, multi-stage printing processes that require precise spatial relationships between sequentially printed components. Notably, once the reference geometries and relative positions were pre-established, the entire alignment process could be accomplished <15 seconds after scanning.

**Shadow correction of pillar occlusions**

An occluding structure comprising ten parallel pillars of 0.5mm diameter attached to a base was fabricated using a Formlabs Form 3B+ stereolithography (SLA) printer with opaque grey resin (Formlabs Gray Resin V4) (Fig. 3c). The occluder was placed into vials and filled with 1mL of 10% GelMA + 0.1% LAP w/v. The GRACE workflow was performed as previously described, but without emission filters during scanning. Instead, the reflected and scattered laser line was imaged over a $\Theta_{total} = 2\pi$ sweep, enabling approximate surface reconstruction by using the light sheet as a profilometer. Cluster detection was used on the surface data to identify the centroids, and a principal component analysis determined the pitch and yaw of the pillars. A parametric model was prepared in Grasshopper to generate the representative occluding to overlay with the scanned data at the correct angles and locations. This 3D model of the occlusion, along with a symmetrical cog model, was exported for processing using the OSMO algorithm[21] to create an optimized set of shadow-corrected projections for printing.

**Shadow correction for ball-in-cage model**

A spherical cage-like occluding structure was fabricated using stereolithography (Fig. 3e), following the same protocol and material as with the occluding pillars. The geometry was algorithmically generated within Grasshopper through a three-step process: i) 45 nodes were randomly distributed on the surface of a 10mm diameter sphere; ii) For each node, paths were computed to its five nearest neighbours based on spatial proximity; iii) Struts of 0.5 mm diameter were generated along these paths, connecting each node to its five nearest neighbours. This procedure yielded a complex, non-uniform, interconnected spherical cage structure, presenting a challenging occlusion scenario for volumetric printing. The cage was embedded within the resin, thermally gelled, and scanned using the light sheet as a profilometer, as previously described. This scan generated a sparse point cloud representing the occluding surface. In contrast to the pillar experiment, where the complete occlusion geometry was reconstructed parametrically, here we instead employed the auto-alignment protocol previously developed for the femur-cartilage model. A reference cage mesh was algorithmically aligned to the scan-derived point cloud, enabling the correctly oriented reference mesh to serve as the computational occlusion volume during projection optimization using the OSMO algorithm. A 5 mm diameter sphere positioned within the cage served as the target geometry for optimization and volumetric printing. Post-printing, the corrected and uncorrected structures (printed in triplicate) were washed, removed from the occluding cages, and imaged using linear light sheet microscopy. Prints were quantified based on their root-mean-squared error and sphericity (Fig. 3f,g,h).

**Analysis of shadow corrected ball-in-cage prints**

To quantifiably compare the corrected and uncorrected spherical prints, the following metrics were used: *i)* root-mean-square error (RMSE); and *ii)* sphericity ($\psi$). To perform this, the resultant light sheet scans of the samples were segmented, and used to reconstruct polygon meshes of the prints. The RMSE is based on the deviation from the ideal target geometry, expressed as:

$$RMSE = \frac{1}{N}\sum_{i=1}^{N} d_i^2 \quad (1)$$

where N is the number of points sampled and $d_i$ is the deviation of that point from the corresponding point on the target geometry. Sphericity is a measure of how closely the object resembles a sphere, and is defined as the ratio of the surface area of a sphere (with the same volume as the object) to the surface area of the object:

$$\psi = \frac{\pi^{\frac{1}{3}}(6V_p)^{\frac{2}{3}}}{A_p} \quad (2)$$

where $V_p$ is the volume, and $A_p$ the surface area of the object.

**iβ-cells subculture and expansion**

iβ-cells, an engineered pancreatic cell line mimicking β-cell function and capable of releasing insulin together with the luminescent reporter NanoLuc, were obtained as previously described in literature[26,27]. iβ-cell were cultured (95% humidified incubator at 37°C, 5% CO2) in Roswell Park Memorial Institute (RMPI) 1640 Medium, containing GlutaMAX™ and HEPES (Gibco, Life technologies) supplemented with foetal bovine serum (FBS, 10% v/v), 100 U/mL penicillin and 100 µg/mL streptomycin. For all experiments, cells were used at passages 3 and 4.

**GRACE printing within constructs produced by embedded extrusion**

A toroidal extrusion path of 4mm in diameter was designed and saved as a G-code file. First, the volumetric printing vial was loaded with 5% GelMA at 37°C + 0.1% w/v LAP. The vial was left to thermally gelate at room temperature overnight, then placed in a R-GEN 100 extrusion printer (RegenHu, Switzerland) and centrally retained using a custom bracket. The GelMA hydrogel was used as a photoreactive suspension bath for embedded extrusion printing. Extrusion printing was performed with the R-GEN 100 printer, via a pneumatic-driven extrusion printhead. As bioink for extruding the toruses, iβ-cells ($5.0 \times 10^7$ cells/mL, stained with the Vybrant DiD membrane dye, ThermoFisher Scientific, to facilitate imaging) were suspended in a 2% w/v alginate solution, used as fugitive viscosity enhancer to improve printing resolution of the high-density cell suspension. The bioink was loaded into 3 mL cartridges equipped with a 23 G stainless steel straight, cylindrical needle (Nordson EFD, USA). Toruses were then extrusion-printed in sterile conditions within the GelMA support bath. After this, the vial containing the support bath and the cell-laden bioink was transferred to the volumetric printer. Following the GRACE workflow, the samples were imaged via light sheet, and a set of blood-vessel mimetic capillary networks were parametrically generated by the software to wrap around the cell-laden toruses (Fig. 4a). Using tomographic volumetric printing, the design was sculpted into the GelMA bath, forming the biological tissue construct. This produced tapered capillaries around the toruses with a 300μm offset, fixed surface area of 180±10 mm², minimum diameter of 450μm, and an inlet and outlet of ~1mm at both ends of a cylindrical bulk scaffold. After volumetrically printing these constructs, the vial was heated to 37°C to dissolve the unpolymerized GelMA and the sample was retrieved and washed with prewarmed PBS. Additionally, two control groups were produced: i) a bulk cylindrical structure containing no channels printed around the extruded toruses as negative control, and ii) a set of randomly gendered channels running along the length of the cylindrical bulk, also maintaining the same 180±10 mm² surface area. Equal surface areas of the random and GRACE-printed constructs were verified post-printing by using linear light sheet to scan the samples, then manually segmenting the channels and determining the surface areas (Fig. 4b,c).

**Analysis of embedded extrusion prints**

The printed constructs (n=6 for each group) with different generated vessel networks were retrieved and cultured overnight in incubator in presence of RPMI 1640 Medium, GlutaMAX™, HEPES (Gibco, Life technologies) supplemented with 10% v/v FBS and 1% penicillin/streptomycin. The following day, the supernatant was collected for each different condition and the bioluminescent reporter nanoLuc (directly related to the amount of insulin stored and released by iβ-cells) was quantified using the NanoLuc Luciferase Kit (Promega Corporation, USA) against a standard curve, using a CLARIOStar Plus multimodal plate reader (BMG Labtech, Germany) (Fig. 4d).

**Osteochondral Differentiation in the cell-laden femur-cartilage model**

At passage 4, ACPCs and MSCs were encapsulated in 10% w/v GelMA solution in PBS at a final density of $1.0 \times 10^7$ cells/mL and $5.0 \times 10^6$ cells/mL, respectively. LAP dissolved in PBS at 0.1% w/v was used as a photoinitiator for the crosslinking. To minimize scattering due to the presence of cells, 30% v/v of Iodixanol (Optiprep™) was used in the MSC containing resin, while a 20% concentration was used for resin containing ACPCs. Vybrant™ DiD and DiO (ThermoFischer Scientific) were used as cell-labelling membrane staining for MSCs and ACPCs, respectively. Cells were stained for 30 min at 37°C according to the manufacturer's instructions. Bone-cartilage models were scanned, aligned, and printed with GRACE (Fig. 4e), in the same fashion as was described previously with acellular constructs. Post-printing, the cell-laden constructs containing both the bone and cartilage layer were washed in warm PBS to remove the unpolymerized biomaterial and cultured in 1:1 ratio osteogenic (Osteogenic Differentiation Medium BulletKit, PT-3002, Lonza) and chondrogenic media (DMEM, supplemented with 100U/mL penicillin, 100µg/mL streptomycin, 0.2mM L-ascorbic acid-2-phosphate, 1% v/v Insulin-Transferring-Selenium (ITS) + Premix Universal Culture Supplement (Corning, USA), $0.1 \times 10^{-6}$ M dexamethasone, and 10ng/mL recombinant human TGF-ß1) for 4 weeks. Media was refreshed 3 times per week.

**Combining FLight with GRACE**

Each stained population of spheroids was scanned using either an excitation wavelength of 450 nm (for DiO) or 650 nm (for DiD), in conjunction with the appropriate emission bandpass filter. The centroid coordinates of each spheroid population were determined and synchronized with a parametric model. This model was designed such as to position either a cylindrical, or star shaped geometry centered at each spheroid, with DiO stained features receiving circles and DiD receiving stars. Following model export and system homing, we employed FLight fabrication (Fig. 4g,h,j). This involved projecting a single binary image of the geometry at its face-on angle into the stationary vial, with the rotational stage fixed at the corresponding angle of the projection. Illumination was maintained until crosslinking was observed through approximately 70% of the vial volume, as determined by real-time schlieren imaging. The un-crosslinked material was then washed in warm PBS, linear light sheet imaged in the vial, then collected and imaged under confocal microscopy (Fig. 4i).

**Statistical analysis**

Data are expressed as mean ± standard deviation (SD), with a minimum sample size of n ≥ 3. Statistical analyses for experiments involving shadow correction were conducted using OriginPro 8.5 (OriginLab, USA), while Prism 9 software (GraphPad Inc., USA) was used for the EmVP experiments. Pairwise comparisons between two groups were performed using a Student's 2-sample t-test (α=0.05), with statistical significance defined as $p < 0.05$. For analyses involving more than two groups, a one-way analysis of variance (ANOVA) was employed with post hoc Turkey's multiple comparison test.

# References

1. Manero, A., Smith, P., Sparkman, J., Dombrowski, M., Courbin, D., Kester, A. & Chi, A. Implementation of 3D printing technology in the field of prosthetics: past, present, and future. *Int. J. Environ. Res. Public Health* **16**, 1641 (2019).
2. Bănică, C.-F., Sover, A. & Anghel, D.-C. Printing the future layer by layer: a comprehensive exploration of additive manufacturing in the era of Industry 4.0. *Appl. Sci.* **14**, 9919 (2024).
3. Culbreath, C. J., Taylor, M. S., McCullen, S. D. & Mefford, O. T. A review of additive manufacturing in tissue engineering and regenerative medicine. *Biomed. Mater. Devices* **2024**, 1–22 (2024).
4. Hoffmann, M. & Elwany, A. In-space additive manufacturing: a review. *J. Manuf. Sci. Eng.* **145**, 020801 (2023).
5. Núñez Bernal, P. *et al.* Volumetric bioprinting of complex living-tissue constructs within seconds. *Adv. Mater.* **31**, 1904209 (2019).
6. Kelly, B. E., Bhattacharya, I., Heidari, H., Shusteff, M., Spadaccini, C. M. & Taylor, H. K. Volumetric additive manufacturing via tomographic reconstruction. *Science* **363**, 1075–1079 (2019).
7. Orth, A. *et al.* On-the-fly 3D metrology of volumetric additive manufacturing. *Addit. Manuf.* **56**, 102869 (2022).
8. Di Cataldo, S., Vinco, S., Urgese, G., Calignano, F., Ficarra, E., Macii, A. & Macii, E. Optimizing quality inspection and control in powder bed metal additive manufacturing: challenges and research directions. *Proc. IEEE* **109**, 326–346 (2021).
9. Rill-García, R., Dokladalova, E., Dokládal, P., Caron, J. F., Mesnil, R., Margerit, P. & Charrier, M. Inline monitoring of 3D concrete printing using computer vision. *Addit. Manuf.* **60**, 103175 (2022).
10. Bisheh, M. N., Chang, S. I. & Lei, S. A layer-by-layer quality monitoring framework for 3D printing. *Comput. Ind. Eng.* **157**, 107314 (2021).
11. Dimas, L. S., Bratzel, G. H., Eylon, I. & Buehler, M. J. Tough composites inspired by mineralized natural materials: computation, 3D printing, and testing. *Adv. Funct. Mater.* **23**, 4629–4638 (2013).
12. Riffe, M. B. *et al.* Multi-material volumetric additive manufacturing of hydrogels using gelatin as a sacrificial network and 3D suspension bath. *Adv. Mater.* **36**, 2309026 (2024).
13. Größbacher, G. *et al.* Volumetric printing across melt electrowritten scaffolds fabricates multi-material living constructs with tunable architecture and mechanics. *Adv. Mater.* **35**, 2300756 (2023).
14. Barbera, L. *et al.* Multimaterial volumetric printing of silica-based glasses. *Adv. Mater. Technol.* **9**, 2202117 (2024).
15. Lee, A. *et al.* 3D bioprinting of collagen to rebuild components of the human heart. *Science* **365**, 482–487 (2019).
16. Liu, H. *et al.* Filamented light (FLight) biofabrication of highly aligned tissue-engineered constructs. *Adv. Mater.* **34**, 2204301 (2022).
17. Ester, M., Kriegel, H.-P., Sander, J. & Xu, X. A density-based algorithm for discovering clusters in large spatial databases with noise. In *Proc. 2nd Int. Conf. Knowledge Discovery and Data Mining (KDD-96)* 226–231 (1996).
18. Chansoria, P. *et al.* Synergizing algorithmic design, photoclick chemistry and multi-material volumetric printing for accelerating complex shape engineering. *Adv. Sci.* **10**, 2300912 (2023).

19. Bagheri, A., Zakerzadeh, M. R. & Sadigh, M. J. Occlusion-based model weighting for volumetric additive manufacturing around inserts. *Virtual Phys. Prototyp.* **19**, e2407473 (2024).
20. Bernal, P. N. N. *et al.* Volumetric bioprinting of organoids and optically tuned hydrogels to build liver-like metabolic biofactories. *Adv. Mater.* **34**, 2110054 (2022).
21. Rackson, C. M. *et al.* Object-space optimization of tomographic reconstructions for additive manufacturing. *Addit. Manuf.* **48**, 102367 (2021).
22. Darkes-Burkey, C. & Shepherd, R. F. Volumetric 3D printing of endoskeletal soft robots. *Adv. Mater.* **36**, 2402217 (2024).
23. Skylar-Scott, M. A. *et al.* Biomanufacturing of organ-specific tissues with high cellular density and embedded vascular channels. *Sci. Adv.* **5**, eaaw2459 (2019).
24. Szklanny, A. A. *et al.* 3D bioprinting of engineered tissue flaps with hierarchical vessel networks (VesselNet) for direct host-to-implant perfusion. *Adv. Mater.* **33**, 2102661 (2021).
25. Grigoryan, B. *et al.* Multivascular networks and functional intravascular topologies within biocompatible hydrogels. *Science* **364**, 458–464 (2019).
26. Ribezzi, D. *et al.* Shaping synthetic multicellular and complex multimaterial tissues via embedded extrusion-volumetric printing of microgels. *Adv. Mater.* **35**, 2301673 (2023).
27. Mansouri, M., Xue, S., Hussherr, M.-D., Strittmatter, T., Camenisch, G. & Fussenegger, M. Smartphone-flashlight-mediated remote control of rapid insulin secretion restores glucose homeostasis in experimental type-1 diabetes. *Small* **17**, 2101939 (2021).
28. Regehly, M. *et al.* Xolography for linear volumetric 3D printing. *Nature* **588**, 620–624 (2020).
29. Kuang, X. *et al.* Self-enhancing sono-inks enable deep-penetration acoustic volumetric printing. *Science* **382**, 1148–1155 (2023).
30. Habibi, M., Foroughi, S., Karamzadeh, V. & Packirisamy, M. Direct sound printing. *Nat. Commun.* **13**, 1800 (2022).
31. Wu, Q. *et al.* Embedded extrusion printing in yield-stress-fluid baths. *Matter* **5**, 3775–3806 (2022).
32. Nguyen, T. L. *et al.* Quantitative phase imaging: recent advances and expanding potential in biomedicine. *ACS Nano* **16**, 11516–11544 (2022).
33. Munck, S. *et al.* Challenges and advances in optical 3D mesoscale imaging. *J. Microsc.* **286**, 201–219 (2022).
34. Kumar, M., Pensia, L. & Kumar, R. Single-shot off-axis digital holographic system with extended field-of-view by using multiplexing method. *Sci. Rep.* **12**, 16462 (2022).
35. Batalov, I. *et al.* Grayscale 4D biomaterial customization at high resolution and scale. *bioRxiv* 2024.01.31.578280 [Preprint] (2024); https://doi.org/10.1101/2024.01.31.578280.
36. Falandt, M. *et al.* Spatial-selective volumetric 4D printing and single-photon grafting of biomolecules within centimeter-scale hydrogels via tomographic manufacturing. *Adv. Mater. Technol.* **8**, 2300026 (2023).
37. Wang, B. *et al.* Stiffness control in dual color tomographic volumetric 3D printing. *Nat. Commun.* **13**, 367 (2022).
38. Carberry, B. J., Rao, V. V. & Anseth, K. S. Phototunable viscoelasticity in hydrogels through thioester exchange. *Ann. Biomed. Eng.* **48**, 2053–2063 (2020).
39. Yein, R. Dendro. *GitHub*; https://github.com/ryein/dendro.
40. Piacentino, G. Shortest Walk Gh. *Food4Rhino*; https://www.food4rhino.com/en/app/shortest-walk-gh.

41. Hart, P. E., Nilsson, N. J. & Raphael, B. A formal basis for the heuristic determination of minimum cost paths. *IEEE Trans. Syst. Sci. Cybern.* **4**, 100–107 (1968).
42. Lim, K. S. *et al.* One-step photoactivation of a dual-functionalized bioink as cell carrier and cartilage-binding glue for chondral regeneration. *Adv. Healthc. Mater.* **9**, 1901792 (2020).

**Acknowledgments:** This work was performed at the Regenerative Medicine Centre Utrecht, the Netherlands. The authors thank Prof. Jason Burdick for the constructive feedback on the manuscript.

**Author contributions:**

    Conceptualization: SF, RL

    Methodology: SF, GG, DR, MG, RL

    Investigation: SF, GG, DR, AL

    Visualization: SF, DR, AL, MG, EG, RL

    Funding acquisition: RL

    Project administration: RL

    Supervision: RL

    Writing – original draft: SF, GG, RL

    Writing – review & editing: SF, GG, DR, AL, MG, JM, RL

**Funding:** This project received funding from the European Research Council (ERC) under the European Union's Horizon 2020 research and innovation program (grant agreement no. 949 806, VOLUME-BIO). R.L and J.M. acknowledge the funding from the Gravitation program "Materials Driven Regeneration" (023.003.013) funded by the Netherlands Organization for Scientific Research (024.003.013). R.L. acknowledges financial support from Dutch Research Council's Talent program (Vidi, 20387).

**Competing interests:** S.F. and R.L. are inventors on a provisional patent application PCT/NL2024/050642 that covers part of the workflow reported in this manuscript. R.L. is scientific advisor for Readily3D SA. The other authors declare no competing interests.

**Supplementary Table 1.** Table of key process durations with the various modalities of GRACE explored within our work.

| | Duration [s] | Additional Info. |
|---|---|---|
| **GRACE Printing** | | |
| Light sheet scan ($\Theta_{total} = \pi$) | 80-120 | Double for $\Theta_{total} = 2\pi$. Dependent on signal exposure time per slice. |
| Feature isolation & coordinate export | 10-30 | Dependent on feature quantity. |
| Parametric model generation & export | 1-10 | Dependent on parametric model. |
| Back-projection generation | 60-80 | Ram-Lak filtered tomographic back projection. |
| Printing | 20-40 | Model and input power dependent. Multiply above by number of different spectral channels used. |
| **GRACE Printing with Sequential Auto-alignment** | | |
| Light sheet scan ($\Theta_{total} = \pi$) | 80-120 | Double for $\Theta_{total} = 2\pi$. Dependent on signal exposure time per slice. |
| Feature isolation & coordinate export of pre-printed geometry. | 10-30 | Dependent on feature quantity. |
| Generation of point cloud within reference model to facilitate alignment with scanned point cloud. | 10-20 | Only performed once and can be pre-calculated. |
| Alignment using iterative closest point algorithm. | 1-5 | Performed in MATLAB using 'pcregistericp' function. |
| Parametric model generation | 1-2 | Transformation of geometry to be printed, onto scanned geometry. |
| Back-projection Generation | 60-80 | Ram-Lak filtered tomographic back projection. |
| Printing | 20-40 | Model and input power dependent. |
| **GRACE Printing with Shadow Correction** | | |
| Light sheet scan ($\Theta_{total} = 2\pi$) | 160-240 | Double for $\Theta_{total} = 2\pi$. Dependent on signal exposure time per slice. |
| Occlusion mapping and export of raw surface point cloud. | 10-30 | Depending on occlusion size or quantity. |
| Parametric generation of occlusion surface (or alignment as above). | 1-10 | Dependent on parametric model. |
| Parametric occlusion model generation & export. | 5-10 | Transformation of geometry to be printed, onto scanned geometry. |
| Optional back-projection optimization of target model with occlusion using OSMO. Step used only in the experiments reported in Figure 3. | <7200 | OSMO Optimization with 20 iterations. |
| Printing | 20-40 | Model and input power dependent. |